\documentclass[conference]{IEEEtran}
\IEEEoverridecommandlockouts
\usepackage[noadjust]{cite}
\usepackage{amsmath,amssymb,amsfonts,amsbsy}
\usepackage{algorithmic}
\usepackage{graphicx}
\usepackage{textcomp}
\usepackage{xcolor}
\usepackage{enumitem}
\usepackage{etoolbox}
\usepackage{multirow}
\usepackage{svg}
\usepackage{arydshln}
\usepackage{mathtools}
\usepackage{subfig}
\usepackage{caption}
\usepackage{subcaption}
\usepackage{amsmath}
\usepackage{cite}

\usepackage{caption,subcaption}
\usepackage{caption}
\usepackage{subcaption}
\usepackage{dblfloatfix}
\usepackage{amsmath,amssymb,amsfonts,amsbsy,amsthm}

\usepackage{array}

\newcolumntype{C}[1]{>{\centering\let\newline\\\arraybackslash\hspace{0pt}}m{#1}}

\usepackage{hyperref}
\usepackage{array}
\newcolumntype{P}[1]{>{\centering\arraybackslash}p{#1}}
\makeatletter
\newcommand*{\rom}[1]{\expandafter\@slowromancap\romannumeral #1@}
\makeatother
\usepackage{amsthm}
\theoremstyle{definition}

\usepackage{dblfloatfix}
\usepackage{tabularx,booktabs}
\newcolumntype{Y}{>{\centering\arraybackslash}X}
\newcolumntype{b}{X}
\newcolumntype{s}{>{\hsize=.75\hsize}X}
\newcolumntype{t}{>{\hsize=1.25\hsize}X}
\def\BibTeX{{\rm B\kern-.05em{\sc i\kern-.025em b}\kern-.08em
    T\kern-.1667em\lower.7ex\hbox{E}\kern-.125emX}}

\usepackage{graphicx,wrapfig}

\usepackage{array}
\newcolumntype{L}{>{\centering\arraybackslash}m{3cm}}
\usepackage{url}
\makeatletter
\g@addto@macro{\UrlBreaks}{\UrlOrds}
\makeatother

\usepackage[singlespacing]{setspace} 
\usepackage{listings}
\usepackage{xcolor}

\usepackage{fontawesome5}
\usepackage{threeparttable}

\begin{document}


\title{
Empirical Assessment of Post-Quantum Cryptography Adoption Status in TLS 
}

\author{\IEEEauthorblockN{\textsuperscript{} Md Ariful Haque\textsuperscript{(\faEnvelope[regular])}, Sander Vinberg, Ben Edwards, Jo\~ao Cruz}
\IEEEauthorblockA{\textit{Bitsight Technologies, Boston, MA, USA} \\
\{mdariful.haque, sander.vinberg, benjamin.edwards, joao.cruz\}@bitsighttech.com}
}

\maketitle

\makeatletter
\newif\if@borderstar
\def\bordermatrix{\@ifnextchar*{%
\@borderstartrue\@bordermatrix@i}{\@borderstarfalse\@bordermatrix@i*}%
}
\def\@bordermatrix@i*{\@ifnextchar[{\@bordermatrix@ii}{\@bordermatrix@ii[()]}}
\def\@bordermatrix@ii[#1]#2{%
\begingroup
\m@th\@tempdima8.75\p@\setbox\z@\vbox{%
\def\cr{\crcr\noalign{\kern 2\p@\global\let\cr\endline }}%
\ialign {$##$\hfil\kern 2\p@\kern\@tempdima & \thinspace %
\hfil $##$\hfil && \quad\hfil $##$\hfil\crcr\omit\strut %
\hfil\crcr\noalign{\kern -\baselineskip}#2\crcr\omit %
\strut\cr}}%
\setbox\tw@\vbox{\unvcopy\z@\global\setbox\@ne\lastbox}%
\setbox\tw@\hbox{\unhbox\@ne\unskip\global\setbox\@ne\lastbox}%
\setbox\tw@\hbox{%
$\kern\wd\@ne\kern -\@tempdima\left\@firstoftwo#1%
\if@borderstar\kern2pt\else\kern -\wd\@ne\fi%
\global\setbox\@ne\vbox{\box\@ne\if@borderstar\else\kern 2\p@\fi}%
\vcenter{\if@borderstar\else\kern -\ht\@ne\fi%
\unvbox\z@\kern-\if@borderstar2\fi\baselineskip}%
\if@borderstar\kern-2\@tempdima\kern2\p@\else\,\fi\right\@secondoftwo#1 $%
}\null \;\vbox{\kern\ht\@ne\box\tw@}%
\endgroup
}
\makeatother

\begin{abstract}


Anticipating Q-day security concerns, major CDNs and cloud providers are migrating to post-quantum cryptography (PQC). While existing literature benchmarks PQC performance, empirical research evaluating real-world adoption data remains scarce. We present an empirical assessment of PQC deployment across seven critical protocols (HTTPS, LDAPS, SMTPS, IMAP-SSL, POP3-SSL, TELNETS, and RDP) using 10,000 target IP addresses per protocol. At the origin-server level, PQC algorithm support reaches 9.1\% for HTTPS and 7.39\% for LDAPS, while falling below 1\% for all other evaluated protocols. At the entity level, PQC support is 11\% for HTTPS and 5.4\% for LDAPS, while the remaining protocols have less than 5\% of entities supporting PQC on the hosted servers. We further correlate these adoption trends against organizational digital and financial footprint, and industry sectors, yielding a multi-dimensional empirical assessment of PQC deployment. Finally, at the domain level, we perform a comparative analysis of PQC adoption in Cloudflare-hosted domains against baseline Cloudflare datasets. Collectively, our findings provide critical insights into the current state of PQC readiness over Transport Layer Security (TLS).
\end{abstract}

\begin{IEEEkeywords}
Post-Quantum Cryptography, Cybersecurity, Encryption Algorithms, TLS1.3, MLKEM, HTTPS, LDAPS, Empirical Analysis.
\end{IEEEkeywords}

%
\section{Introduction}
Transport Layer Security (TLS) relies on classical public-key algorithms (such as Diffie-Hellman, Elliptic-Curve cryptography, or RSA) that will eventually be broken by sufficiently powerful quantum computers \cite{shor1999polynomial}. To prevent attackers from exploiting ``harvest-now, decrypt-later (HNDL)'', large content delivery networks (CDNs) are already adopting PQC. In October 2025, Cloudflare claimed that \textit{``over 50\% (and rising!) of human traffic is protected against store-now/decrypt-later, making post-quantum key agreement the new security baseline for the Web''} \cite{westerbaan2025pqc}. Also, Google has announced a 2029 timeline to secure the quantum era with post-quantum cryptography migration \cite{adkins2026pqc}. Thus, it is high time to conduct a data-driven empirical analysis of PQC readiness across TLS to assess PQC adoption status and trends.

For the assessment of PQC readiness, we focus specifically on the TLS protocol due to its ubiquitous role in securing modern network communications and information exchange. Crucially, PQC algorithm support is exclusive to TLS version 1.3, as legacy versions (e.g., TLS 1.1 and TLS 1.2) lack the native architecture required for post-quantum key encapsulation mechanisms. In this work, we investigate the real-world PQC adoption status across seven prominent secure services running over TLS, categorized by their operational domains: web traffic via HTTPS (port 443); directory services via LDAPS (port 636); mail delivery and retrieval infrastructure via SMTPS (port 465), IMAP-SSL (port 993), and POP3-SSL (port 995); and remote administration protocols via TELNETS (port 992) and RDP (port 3389). 

In the current state of the art, several studies highlight the need for PQC readiness in enterprise networks \cite{qureshi6045163risk, kundu2025quantum}. In a recent study, Mallick et al. \cite{mallick2026studypostquantumstatus} surveyed nine widely deployed protocols for PQC adoption feasibility — TLS, IPsec, BGP, DNSSEC, SSH, QUIC, OpenID Connect, OpenVPN, and the Signal Protocol. For each protocol, they provided cryptographic reliance, post-quantum migration efforts to date, and the challenges that still remain. Demir et al. \cite{demir2025performance} provide a performance study of CRYSTALS-Kyber and CRYSTALS-Dilithium by benchmarking execution times across cryptographic operations such as key generation, encapsulation, decapsulation, signing, and verification, and assessing hardware acceleration benefits. Hanna et al. \cite{hanna2025comprehensive} analyzed the viability of Post-Quantum algorithms in TLS to power-constrained IoT devices, focusing on two widely used IoT network protocol stacks (Bluetooth Low Energy (BLE) and Wi-Fi). Additionally, Sowa et al. \cite{sowa2024post} present PQC adoption rates on TLS and SSH, but the study seems outdated, as it only presents the occurrence of the SNTRUP761 algorithm and does not present the occurrences of the MLKEM algorithm, which is the NIST standard PQC algorithm and is mostly used in TLS. 


These researches focus on operational challenges and performance evaluation of the PQC algorithms, but there is very limited research on the status of real-world PQC adoption. In this work, we present the PQC adoption status on seven secure protocols mentioned above at the server-level as well as entity-level. Within the scope of this report, we looked into several research aspects: 1) PQC adoption at the level of both origin servers as well as entities for each of these protocols, 2) PQC adoption at the domain level, and 3) Variation in PQC support by organizations' industry, revenue, employee size, and digital size (e.g., IPv4 count, etc.). To note here, we use the term `origin server' or `host' to represent a unique IP address that hosts the specified service, and the term `entity' to represent an organization that hosts that IP address. Sometimes we use the term `service' instead of `protocol' interchangeably for the above seven specified protocols, as these are also considered services. The significant contributions of this work are as follows:
\begin{enumerate}[label=\roman*)]
  \item Discussion on PQC algorithms fingerprinting process on TLS, 
  \item Origin server-level and domain-level PQC adoption assessment, along with comparative analysis, and 
  \item Entity-level PQC adoption status based on several  digital size and financial metrics, as well as industry sector. Thus, generating key insights on the current PQC adoption trends. 
\end{enumerate} 

We organize the rest of the paper as follows. Section \ref{sec:2} discusses the data collection and processing methods of PQC algorithm for the specified services. Section \ref{sec:3} is the host-level PQC adoption analysis, followed by the entity-level PQC adoption analysis in Section \ref{sec:4}. Section \ref{sec:5} provides comparative analysis on Cloudflare-hosted IPs and domains. Finally, Section \ref{sec:6} concludes the paper.

\section{Data Collection \& Processing}\label{sec:2}
In this section, we present the data collection and processing methodologies of PQC key exchange (KEX) algorithms pertinent to the selected protocols in TLS. We provide the details below. 

\subsection{Scan Target List} \label{sec:2A}
Bitsight has built an in-house internet scanning platform called GROMA that maintains a continuous global survey of the public-facing Internet. More details are available here \cite{bitsight2024groma}. We use GROMA data to create a list of 10,000 IPv4 addresses per port for the seven selected services to conduct a limited scan. We then utilize OpenSSL (v3.5) for fingerprinting the TLS handshake messages and use an in-house developed tool to perform a large-scale scan on the selected target list. We use an authorized GCP (Google Cloud Platform) instance to perform the scan. 

\subsection{KEX Algorithm Extraction From TLS Handshake} \label{sec:2B}
Several resources \cite{morrissey2008modular, CloudflareTLSHandshake} explained the TLS handshake processes in detail. Essentially, the core of the TLS handshake is the `ClientHello' and `ServerHello' messages, where at first the client initiates the negotiation process by sending a message to the server with its supported TLS versions, a client random number, and a list of supported cipher suites or cryptographic algorithms. The server then responds with its chosen TLS version, a server random number, a digital certificate, and the specific cipher suite it selected. The response message sent by the server includes the key exchange algorithm it selected for the specific TLS communication. We extract the negotiated algorithm from the TLS handshake message traces, incorporating custom-made RegEx expressions in our extraction logic. The extraction process looks into phrases like \textit{``Negotiated TLS1.3 group''} and \textit{``NamedGroup''} within the ``ServerHello'' message and captures the associated algorithm from the TLS handshake message traces.
\subsection{Entity Mapping and Firmographic Data} \label{sec:2C}
Next, we map the scanned IP addresses to entities to perform entity-level analyses. The entity mapping process maps the entity IDs corresponding to the IP addresses or domains, where each entity is a real-world organization. The mapping process relies on the use of an AI-powered asset mapping and attribution engine called `GIA'. More details on the entity mapping process and GIA are available here \cite{bitsight2024gia}. We collect firmographic data (i.e., revenue and employee count) from our service providers. We also leverage the digital metrics data (e.g., IPv4 count and distinct TLS-enabled host count known as 'SSL\_Active\_Count') from different internal data pipelines.

Next, we discuss the main analyses on the assessment of PQC adoption status on the selected protocols. We divide the discussion into two broad sections: server-level analysis \ref{sec:3} and entity-level analysis \ref{sec:4}. 

\section{Server-Level PQC Readiness Analysis}\label{sec:3}

This section provides details of the server-level PQC status. We divide the analyses into several subsections: 1) PQC KEX proportions per service, and 2) Proportions of all observed KEX algorithm variants.

\subsection{PQC KEX Proportions Per Service} \label{subsec:3B}
Out of the total 70,000 scanned IPv4 targets, there are a total of 5,517 (7.89\%) connections that were unsuccessful. The unsuccessful connections are mainly because of two main reasons: \textit{connection refused by the server}, and \textit{connection timed out}. We exclude the unsuccessful connections, and all the subsequent analyses are based on the successful connections. Table \ref{table2} presents the proportions of PQC and Non-PQC supported IPs during TLS1.3 handshake per service for the successful connections only. We observed that HTTPS has the highest number of PQC-supported IPs, which is 816 out of 8960, meaning 9.1\% of the HTTPS target IPs were PQC supported. The same proportion of PQC-supported IPs for the other services is as follows: LDAPS (7.4\%), IMAP-SSL (0.64\%), POP3-SSL (0.51\%), SMTPS (0.45\%), TELNETS (0.15\%), and RDP (0.01\%). Thus, except for HTTPS and LDAPS, we see that a very few of those scanned IPs were supporting PQC, specifically for TELNETS (13 out of 8486 successful connections) and RDP (1 out of 8602 successful connections). 

\begin{table}[!b]
    \centering
    \begin{threeparttable}
        \caption{Proportions of IPs supporting PQC and Non-PQC algorithm during TLS handshake.}
        \label{table2}
        \small
        \newcolumntype{C}{>{\centering\arraybackslash}X}
        
        \begin{tabularx}{\columnwidth}{>{\hsize=0.6\hsize}X >{\hsize=1.2\hsize}C >{\hsize=1.2\hsize}C} 
            \toprule
            \textbf{Service} & \textbf{PQC (\%)} & \textbf{Non-PQC (\%)} \\ 
            \midrule
            HTTPS          & 9.11        & 90.89  \\ 
        LDAPS              & 7.39        & 92.61 \\
        IMAP-SSL           & 0.64            & 99.36 \\
        POP3-SSL           & 0.51        & 99.49 \\
        SMTPS              & 0.45        & 99.55 \\
        TELNETS            & 0.15        & 99.85 \\
        RDP                & 0.01        & 99.99 \\
            \bottomrule
        \end{tabularx}
        
        \begin{tablenotes}
            \small
            \item[*] percentages are out of successful connections per port.
        \end{tablenotes}
    \end{threeparttable}
\end{table}

\begin{table}[!t]
    \centering
    \caption{Observed Server-Level PQC KEX Algorithm Variants in TLS Handshake (All the Selected Ports Combined).}
    \label{table3}
    \small 
    \addtolength{\tabcolsep}{3pt} 
    
    \begin{tabularx}{\columnwidth}{Xrr} 
        \toprule
        \textbf{KEX Algorithm Variants} & \textbf{Count} &\textbf{Percentage (\%)} \\ 
        \midrule
        \texttt{ecdh\_x25519}        & 31,807    &49.33 \\ 
        \texttt{secp256r1}           & 2,465  & 3.82 \\ 
        \texttt{X25519MLKEM768}      & 1,674  & 2.60 \\ 
        \texttt{secp384r1}           & 476  & 0.74 \\ 
        \texttt{secp521r1}           & 204  & 0.32 \\ 
        \texttt{ffdhe3072}           & 2  & 0.0031 \\
        \texttt{TLS1.3 Not Supported} & 27,855 & 43.18\\
        \bottomrule
    \end{tabularx}
\end{table}

\begin{figure*}[!b] 
    \centering
    \includegraphics[width=\textwidth, trim=0.0in 0 0.2in 0, clip]{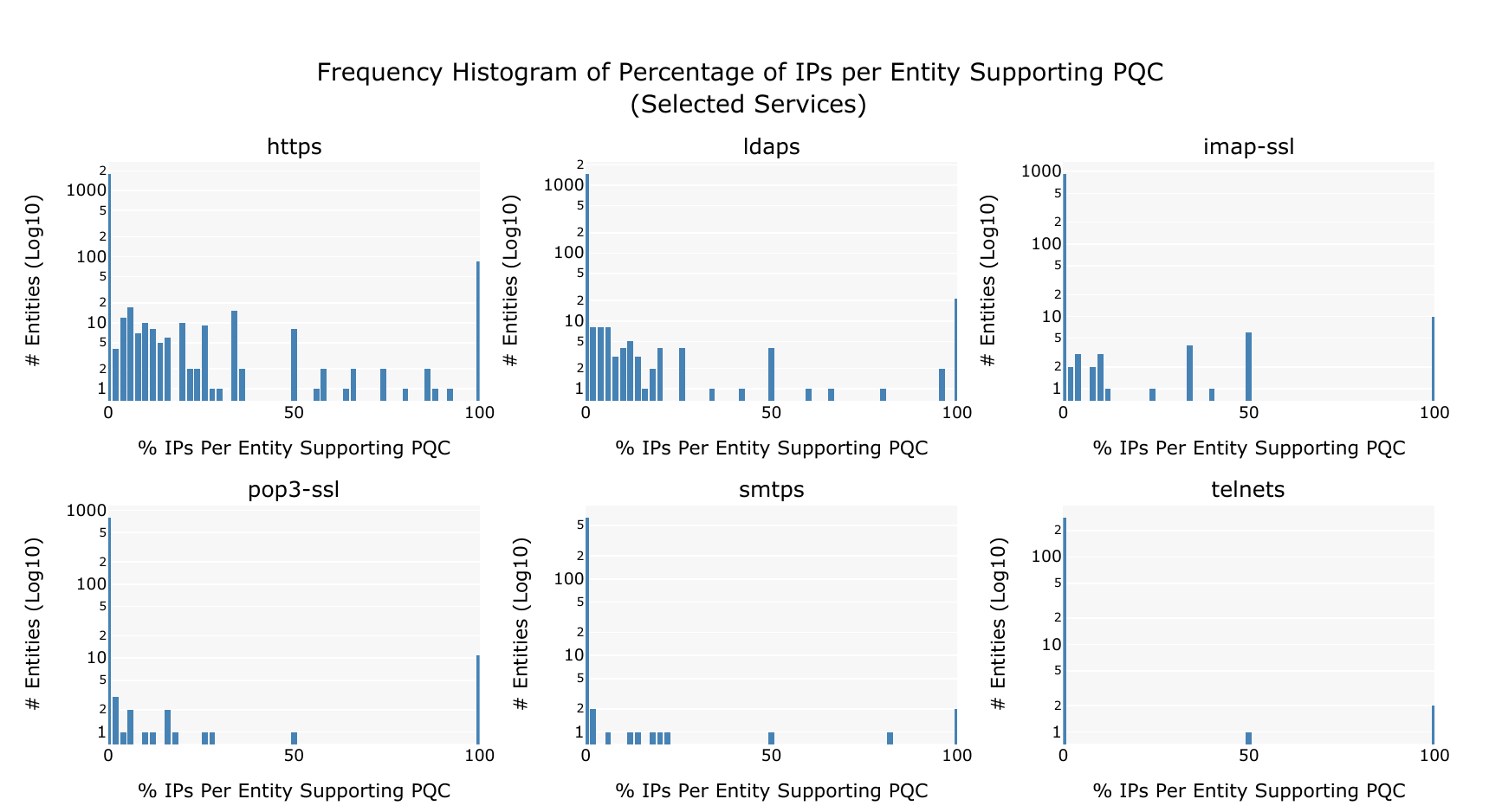}
    \caption{Frequency histogram of percentage of PQC support per entity (selected services).}
    \label{fig:1}
\end{figure*}

\subsection{Proportions of all the observed key exchange algorithms}\label{subsec:3C}

Table \ref{table3} presents the proportion of all the different variants of key exchange algorithms, including both PQC and non-PQC variants, that we observed across all successful TLS1.3 handshake messages. The most dominant key share algorithm is the classical \textbf{ecdh\_x25519} algorithm, appearing around 49.3\% of the total successful scans. The only PQC algorithm observed is hybrid \textbf{X25519MLKEM768}, which appears 2.6\% of the total number of successful connections (across all seven services). The other observed key share algorithms are \textbf{secp256r1}, \textbf{secp384r1}, and \textbf{secp521r1}, and \textbf{ffdhe3072}. The \textbf{ffdhe3072} algorithm was found in two connections only. The remaining cases (43.2\%) were returning \textit{`\textless{NULL}\textgreater'} in the algorithm field (means TLS1.3 not supported).

\section{Entity-Level PQC Adoption Analysis}\label{sec:4}
As illustrated earlier, we perform an entity-join technique to map the scanned servers' IP addresses to specific entities and perform a deduplication process to reduce any duplicate count on the entity-level. 

\begin{table}[!t]
    \centering
    \caption{Proportions of Entities Supporting PQC KEX on 1 or More IPs Per Service.}
    \label{table4}
    \small
    \newcolumntype{C}{>{\centering\arraybackslash}X}
    
    \begin{tabularx}{\columnwidth}{>{\hsize=0.6\hsize}X >{\hsize=1.2\hsize}C >{\hsize=1.2\hsize}C} 
        \toprule
        \textbf{Service} & \textbf{PQC (\%)} & \textbf{Non-PQC (\%)} \\ 
        \midrule
        HTTPS            & 11.04        & 88.96  \\ 
        LDAPS            & 5.42         & 94.58 \\ 
        IMAP-SSL         & 4.00        & 96.00 \\ 
        POP3-SSL         & 3.03         & 96.97 \\ 
        SMTPS            & 1.85        & 98.15 \\ 
        TELNETS          & 1.75        & 98.25 \\ 
        RDP              & 0.11        & 99.89 \\ 
        \bottomrule
    \end{tabularx}
\end{table}

\begin{figure*}[!b] 
    \centering
    \includegraphics[width=0.9\textwidth, trim=0.25in 0 0.25in 0, clip]{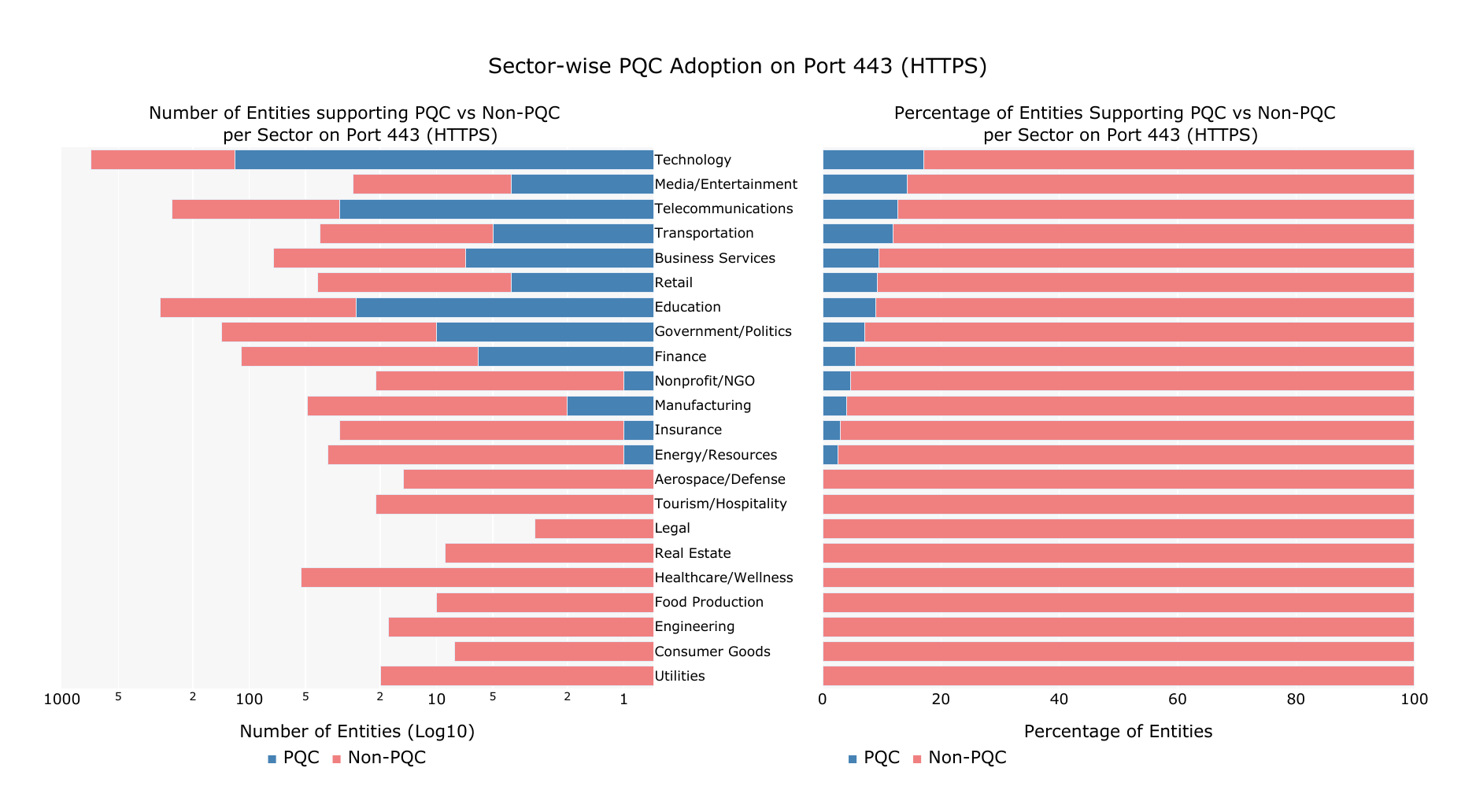}
    \caption{Entities sector-wise PQC adoption on port 443 (HTTPS servers).}
    \label{fig:2}
\end{figure*}

\subsection{Entity-Level PQC Proportions} \label{subsec:4A}
In the collected data, we have 4,636 unique entities; among them, only 308 entities support PQC in 1 or more of their TLS-enabled IPs, and 4,328 entities do not support any of the PQC algorithm variants on those selected ports. The percentage of entities supporting the PQC algorithm (\textbf{X25519MLKEM768}) in one or more IPs per service during TLS handshake is shown in Table \ref{table4}. In the service level, we see HTTPS has $\sim$11\% entities supporting PQC, whereas LDAPS has around 5.4\% PQC-supported entities. For services like IMAP-SSL and POP3-SSL, the percentage of entities supporting PQC is 4\% and 3\%, respectively. For the other services such as SMTPS, TELNETS, and RDP, this percentage of entities supporting PQC drops to \textless{2\%}. Thus, our findings show that while HTTPS has \textgreater{10\%} entities supporting PQC in one or more IPs, this proportion drops to $\sim$5\% or less for the rest of the selected services. 

As the number of IPs controlled by an entity may vary widely, the next research question that we explored is ``what percentage of IPs per entity per service support PQC''. Figure \ref{fig:1} presents an analysis of the percentage of IPs per entity per service supporting PQC. This histogram provides a high-level view of how PQC adoption is distributed across different entities (i.e., organizations). For each graph, the X-axis represents the percentages of IPs per entity supporting PQC, and the Y-axis represents the number of entities on a Log10 scale. 

From this analysis, we see across all services, there is a huge spike at the far left with \textless{1\%} of IPs per entity supporting PQC on its TLS1.3 communication. This indicates that the vast majority of entities still have near-zero PQC support currently. There is a smaller, distinct spike at 100\% in almost every protocol. This suggests a group of entities (likely security-concerned) that have converted their entire infrastructure to support PQC. Compared to legacy protocols like TELNETS or SMTPS, the HTTPS histogram shows much more activity in the middle ranges (10\% - 40\% of PQC support). This indicates that web traffic is seeing a more gradual, phased rollout of PQC than other services. LDAPS is the second service showing a similar growing PQC adoption pattern with some fewer numbers compared to HTTPS. Protocols like SMTPS and TELNETS show almost no partial adoption. 

\subsection{Sector-Wise PQC Adoption} \label{subsec:4B}
Post entity-join, we have in total 22 sectors associated with the entities in the scanned data for the selected services. Here we provide the sector-wise analysis for HTTPS only as HTTPS have \textgreater{10\%} of PQC support in TLS1.3 handshake messages. Figure \ref{fig:2} illustrates the PQC adoption in different sectors for HTTPS (port 443) in mirrored style where it contrasts the absolute volume of PQC adoption by the entities per sector on the left with the relative maturity of each sector on the right. From the figure, we find that the Technology sector is the clear frontrunner, boasting both the highest absolute number of PQC-supporting entities and the highest adoption rate ($\sim$17\%). The Media/Entertainment, Telecommunications and Transportation sectors show higher proportional PQC adoption rates (above 10\%). A significant number of sectors - including Legal, Real Estate, Healthcare, and Utilities show near-zero PQC adoption. These sectors remain almost entirely reliant on classical cryptography. Even in the leading sectors, the vast segment of the bars are non-PQC. This indicates that while PQC is entering the `early adopter' phase in tech-heavy industries, the global transition to quantum-resistant HTTPS (origin-server level) is still in its infancy.

\begin{figure*}[!b] 
    \centering
    \includegraphics[width=0.85\textwidth, trim=0in 0 0.15in 0, clip]{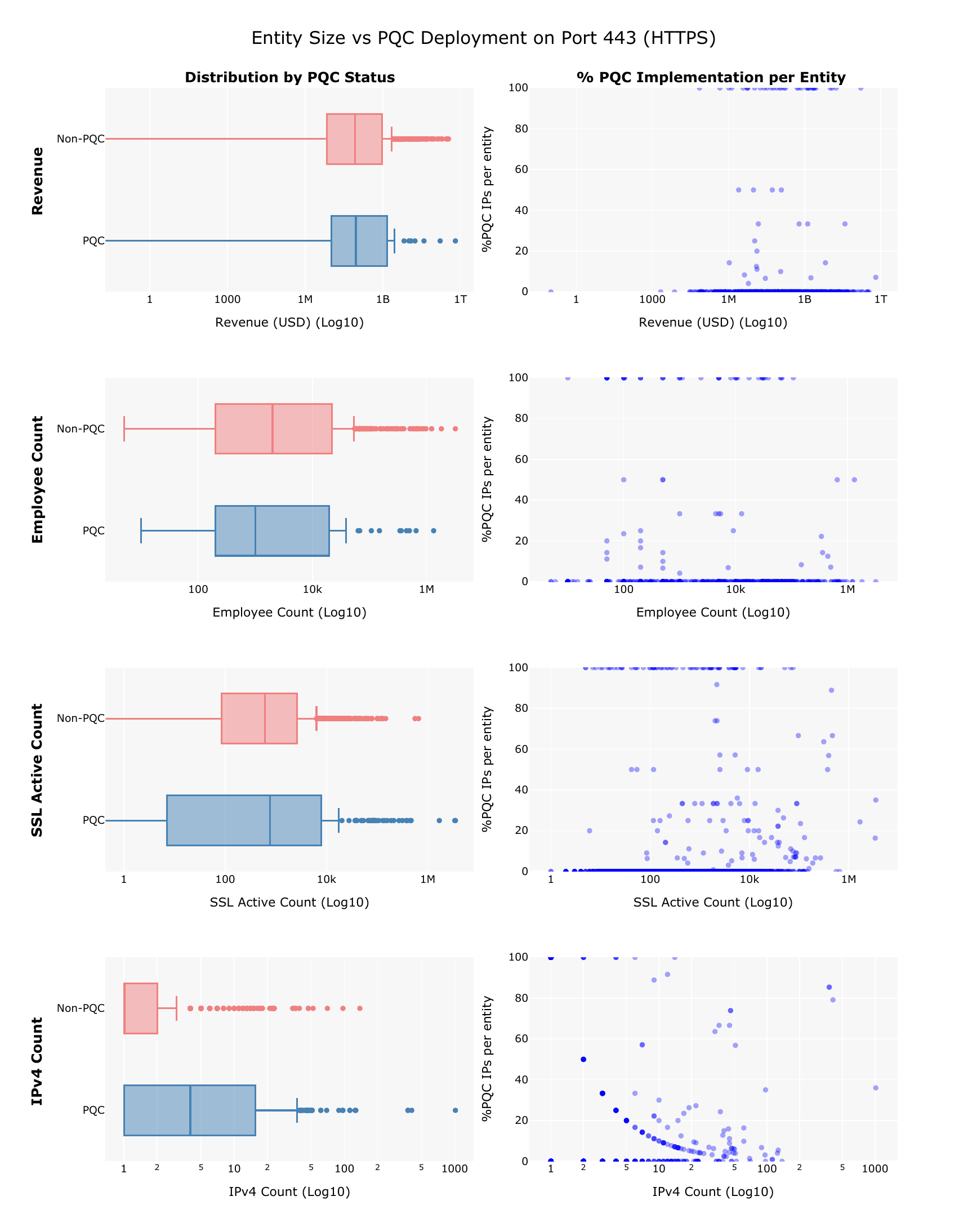}
    \caption{Entities financial and digital size metrics-based PQC adoption in HTTPS servers (Port 443).}
    \label{fig:5}
\end{figure*}

\subsection{Entity Size-Based PQC Adoption} \label{subsec:4C}

We perform entity size-based analysis to understand the relationship between entity size metrics and PQC adoption on TLS, specifically on HTTPS as HTTPS has \textgreater{10\%} of entities supporting PQC on one or more IPs. We selected two broad classes of entity size metrics: financial metrics and digital size metrics. The financial metrics include two sub-metrics: the organization's revenue (in USD) and the number of employees. The digital size metric includes `SSL\_ACTIVE\_Count' and IPv4 address count. To give some context, \textbf{`SSL\_ACTIVE\_Count'} is basically the number of unique SSL evidence keys (i.e., unique findings) graded over the past 60 days. An evidence key is either an `IP:PORT' or `Hostname:Port' information. 

We present the analysis in a 4x2 grid view of the selected entity size metrics in Figure \ref{fig:5} where each row represents a metric, either financial or digital size of the entities. The row metrics are revenue in USD, employee count, SSL active count, and IPv4 count, respectively. The first column of each row indicates the Boxplot distribution of PQC and non-PQC entities over those size metrics. The second column represents the percentage of IPs per entity with PQC adoption vs the entity size metrics. Thus, the first column compares the distribution of PQC-supported entities with that of non-PQC-supported entities. And the second column shows the adoption depth of PQC entities with the percentage of IPs per entity supporting PQC. The X-axes are in Log10 scale because of the large range of those metrics. 

From the Boxplots, PQC adopters have slightly higher median revenues and employee counts, indicating that larger budgets and dedicated security teams drive early adoption. PQC adopters show a noticeably higher median IPv4 count, confirming that organizations with larger network infrastructure are leading the transition. Despite large enterprise leadership, significant overlap in active SSL counts shows that smaller entities are also adopting PQC.

From the scatter plots of PQC deployment density, we find that PQC deployment is currently an ``all-or-nothing'' decision; most entities sit at either 0\% or 100\% implementation, leaving a "hollow middle" (20\% – 80\%). Financial scale (revenue/employees) does not dictate deployment density; once committed, small entities are just as likely to achieve 100\% PQC compliance as large ones. As IPv4 counts scale up, 100\% completion rates drop slightly, suggesting that massive network footprints make full PQC migration more difficult to achieve.

\section{PQC Adoption Rate for Cloudflare-Hosted Domains and IPs} \label{sec:5}
Cloudflare continuously updates its PQC adoption across both hosted domains (HTTPS requests) and origin servers \cite{CloudflareRadarPQC}. At the time of writing this paper, the PQC adoption rate for HTTPS requests is 57.3\% (human traffic) and 7.4\% for origin-servers. We conducted a similar analysis for only Cloudflare-hosted IP addresses and domains. For domains (HTTPS requests), we collect 11,979 distinct Cloudflare-hosted top-level domains from GROMA, and specify the server name identification flag `-servername' in the openssl command. We find that out of 11,979 domain scans, 9,297 times it appeared the PQC algorithm (i.e., X25519MLKEM768), 1070 times ecdh\_x25519, and for the remaining cases, a few other algorithms, including \textit{`\textless{NULL}\textgreater'}. Thus, at the domain-level, we find $\sim$83.6\% of the Cloudflare-hosted domains (HTTPS requests) are PQC enabled, whereas 16.4\% were non-PQC domains. We perform a second scan on 11048 unique Cloudflare-hosted IPs, and we find the PQC algorithm (X25519MLKEM768) in 437 target IPs , \textit{`\textless{NULL}\textgreater'} in 10385 IPs, and ecdh\_x25519 in 3 target IPs. Thus, the percentages of Cloudflare-hosted origin-server IPs supporting PQC are roughly $\sim$4\%, and non-PQC is $\sim$96\%. This is somewhat similar to what Cloudflare Radar is presenting \cite{CloudflareRadarPQC}.

\section{Conclusions}\label{sec:6}
This work presents a comprehensive, multi-dimensional analysis of post-quantum cryptography (PQC) adoption over TLS across seven critical network protocols. Using a localized scanning methodology targeting 10,000 IP addresses per service, our empirical findings reveal that while HTTPS consistently exhibits an origin-server PQC adoption rate of approximately 9\%, deployment across the remaining six protocols remains critically low at less than 5\%. Our organizational analysis uncovers a nuanced relationship between entity scale and PQC readiness: while larger enterprises leverage expansive budgets and dedicated engineering teams to initiate deployment, the inherent complexity of massive network infrastructures introduces significant friction to achieving a 100\% PQC rollout. Consequently, our analysis captures an active, gradual rollout phase wherein organizations systematically deploy PQC across isolated, highly specific subsets of their total IP space rather than executing immediate enterprise-wide migrations. Finally, by cross-validating our domain-level data against public Cloudflare Radar telemetry, we confirm the baseline consistency and representativeness of our findings. Ultimately, this study provides a vital empirical benchmark for early-stage PQC readiness in TLS, highlighting both infrastructure constraints and organizational hurdles that must be cleared before realizing a fully post-quantum-secure internet.




\bibliographystyle{IEEEtran}
\bibliography{reference}





\end{document}